\documentclass[iop,apj,tighten]{emulateapj}
\usepackage{apjfonts} 
\usepackage{amsmath,amstext}
\usepackage{natbib}
\usepackage{nicefrac}
\usepackage[breaklinks,colorlinks,citecolor=blue,linkcolor=magenta]{hyperref} 
\usepackage[all]{hypcap} 
\usepackage{xcolor}

\newcommand{\sgra}{Sgr~A$^*$}

\begin{document}
\title{A Plasmoid model for the Sgr A* Flares Observed With Gravity and CHANDRA}
\author{David Ball,\altaffilmark{1,2} Feryal \"Ozel,\altaffilmark{1}  Pierre Christian,\altaffilmark{1} Chi-Kwan Chan\altaffilmark{1,3} \& Dimitrios Psaltis\altaffilmark{1}}
\altaffiltext{1}{Department of Astronomy and Steward Observatory, Univ. of Arizona, 833 N. Cherry Avenue, Tucson, AZ 85721, USA}
\altaffiltext{2}{Email: davidrball@email.arizona.edu}
\altaffiltext{3}{Data Science Institute, University of Arizona, 1230 N. Cherry Ave., Tucson, AZ 85721, USA}
\begin{abstract}
The Galactic Center black hole \sgra\ shows significant variability and flares in the submillimeter, infrared, and X-ray wavelengths. Owing to its exquisite resolution in the IR bands, the GRAVITY experiment for the first time spatially resolved the locations of three flares and showed that a bright region moves in ellipse-like trajectories close to but offset from the black hole over the course of each event. We present a model for plasmoids that form during reconnection events and orbit in the coronal region around a black hole to explain these observations. We utilize general-relativistic radiative transfer calculations that include effects from finite light travel time, plasmoid motion, particle acceleration, and synchrotron cooling and obtain a rich structure in the flare lightcurves. This model can naturally account for the observed motion of the bright regions observed by the GRAVITY experiment and the offset between the center of the centroid motion and the position of the black hole. It also explains why some flares may be double-peaked while others have only a single peak and uncovers a correlation between the structure in the lightcurve and the location of the flare. Finally, we make predictions for future observations of flares from the inner accretion flow of \sgra\ that will provide a test of this model.
    
\end{abstract}

\maketitle

\section{Introduction}
The low luminosity and the broadband spectrum of the supermassive black hole at the center of the Milky Way, Sagittarius~A$^*$ (\sgra), is thought to arise from a high-temperature, low-density, collisionless, and radiatively inefficient accretion flow (see \citealt{yuan2014} for a review). Long-term monitoring has revealed significant multiwavelength variability and flaring behavior from the submillimeter (\citealt{marrone2008}) to the infrared (\citealt{genzel2003, ghez2005, do2019}) and X-ray wavelengths (\citealt{neilsen2013,haggard2019}). The timescales, polarization measurements, and spectra of these observations suggest that the flares most likely originate in the inner accretion flow from compact magnetized structures emitting synchrotron radiation (\citealt{eckart2006, dodds-eden2009,nishiyama2009,ball2016,ponti2017}). When interpreted in the context of the radiatively inefficient accretion models, these flares offer unique insight into particle acceleration and heating mechanisms in collisionless plasmas, which are fundamental plasma physics processes that are largely unconstrained by laboratory experiments.

Numerous theoretical studies have attempted to explain the observed  flaring  and  variability  of \sgra,  often  invoking transient  structures or the episodic release of energy in compact regions of accretion flows. Some models explored hot spots orbiting in the equatorial plane (\citealt{broderick2005}) or along the jet of a black hole (\citealt{younsi2015}). Others discussed plasma instabilities that cause buoyant magnetic bubbles to rise in the accretion disk, eventually erupting into the corona and forming a current sheet where reconnection may occur (\citealt{li2017}). Indeed, magnetic reconnection has been recognized as potentially playing an important role in the observed variability of \sgra\ and other low-luminosity accretion flows, leading to localized and episodic energy release (\citealt{galeev1979, dodds-eden2010, ball2016, li2017, ball2017}). Various magnetohydrodynamic (MHD) simulations have incorporated this effect (e.g., \citealt{dodds-eden2010,chan2015a,ball2016}). Coupling general-relativistic radiative transfer calculations to general relativistic MHD simulations, these latter studies have found that intermittent magnetized structures (or, "flux tubes'') that copiously radiate synchrotron emission, coupled to the strong gravitational lensing when one of these structures passes behind the black hole, can cause significant IR and X-ray variability and flaring behavior.

Recent observations have revealed additional properties of these multiwavelength flares that are not easy to account for in simple models. Using the {\em Chandra} X-ray Observatory, \citet{haggard2019} characterized the time evolution and spectra of very bright X-ray flares, one of which shows a distinct double peak in its lightcurve, with a time separation between the two peaks of about $\sim 40$~min. Observations with the GRAVITY interferometer on the Very Large Telescope measured astrometrically the motions of the centroids of the emission during three IR flares, which were within 100$\;\mu$as of the black hole (\citealt{gravity2018}; hereafter, "the GRAVITY paper"). One of these flares also showed a distinct double-peaked lightcurve with a timescale similar to that of the Chandra flares. The GRAVITY paper (see also  \citealt{baubock2020}) interpreted the astrometric excursions as orbital motions centered around the black hole, although the central positions of the inferred orbits are different among the three flares and offset from the location of the central black hole. For a black hole mass of $M=4.1 \times 10^6 M_\odot$ at a distance of 8.1\;kpc (e.g., \citealt{abuter2019}), 100$\;\mu$as corresponds to 10 $R_{\rm S}$ and 40 minutes corresponds to the orbital period at 3.5~$R_{\rm S}$, where $R_{\rm S} \equiv 2GM/c^2$ is the Schwarzschild radius. 

In this paper, we show that emission from hot plasmoids orbiting in the funnel region of a black hole accretion flow can account for several previously unexplained aspects of the flares. Plasmoids, which are compact structures of magnetized plasma that are formed from the collapse of a current sheet and contain heated and accelerated particles, are a generic byproduct of reconnection events and are a natural way to explain the presence of hot, compact emitting regions that are offset from the black hole. Plasmoids hierarchically merge and may eventually coalesce into an astrophysically large structure (such as  ``monster'' plasmoids in \citealt{uzdensky2010, giannios2013}).  Because these large plasmoids are highly magnetized and contain all of the high-energy electrons accelerated in the reconnection event, they will radiate copiously as the high-energy electrons cool via synchrotron radiation. Furthermore, as these regions can occur away from the equatorial plane, when viewed from an inclined angle with respect to the black hole spin axis, the observed center of emission will trace out ellipse-like shapes, with their centers offset from the position of the black hole. Finally, because they are prevalent in regions containing low-$\beta$ plasma (where $\beta \equiv kT_e/B^2$ is the ratio of of gas pressure to magnetic pressure), such as in the innermost accretion flow and jet/coronal regions around a black hole, the positions of the centroids will naturally have two preferred directions along an axis: either aligned or anti-aligned with the black hole spin.  This has the potential to explain the aligned trajectories of the flares as well as the differences observed between them, as we will discuss in the next section.

Making use of these characteristics, we construct a plasmoid model orbiting in the jet or coronal region of a black hole, with properties informed by microphysical studies of reconnection and incorporating cooling via synchrotron emission. We include the physics of finite light-travel time, which we show can have a significant effect on both the observed lightcurves and the centroid motion.  We show that this model can not only explain the differences in the orientations between different flares but also the connection between the orientation of the flare's trajectory and the structure in its lightcurve. 

In \S2, we discuss the characteristics of the flares observed with GRAVITY. In \S3, we introduce the formalism for the orbital motion and the energetics of plasmoids. In \S4 and 5, we present trajectories and lightcurves for the two possible orientations of the plasmoid motion with respect to the black hole axis. In \S6, we explore the correlations between the various properties of the flares expected in the plasmoid model and conclude in \S7 with testable predictions and outlook for future observations.  

\begin{figure}[t]
    \centering
    \includegraphics[width=\linewidth]{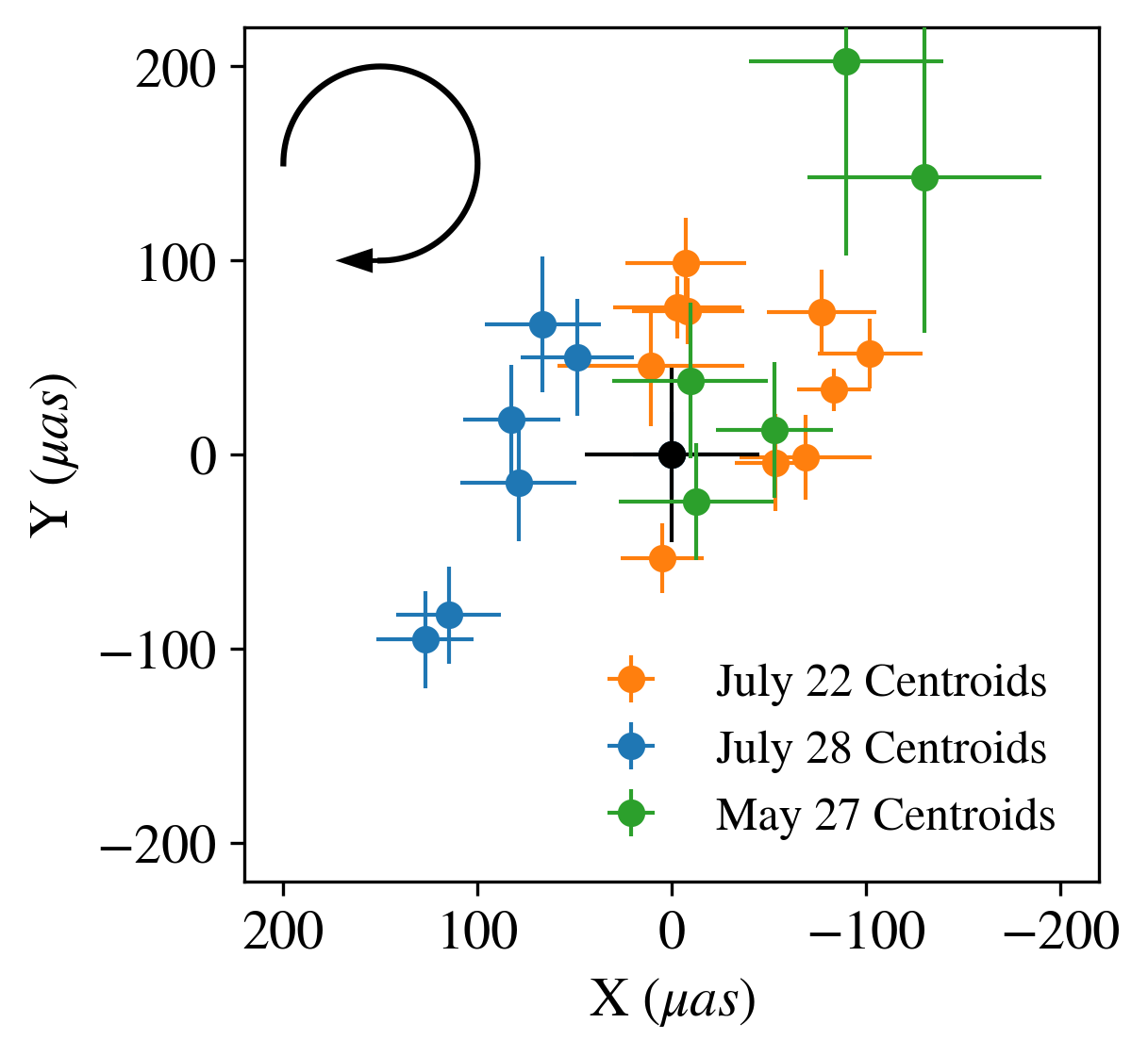}
    \caption{Centroid positions from all three flares observed with GRAVITY, centered on the position of Sgr A* (black point).  The May 27th and July 22nd flares both appear to be similar in their orbital orientation, while the July 28th flare centroids largely point in the opposite direction.  None of the centers of the projected motion are centered on the position of \sgra.  We depict the direction of orbital motion with a black arrow in the upper left hand corner.}
    \label{fig:centroid_data}
\end{figure}

\begin{figure}[t]
    \centering
    \includegraphics[width=\linewidth]{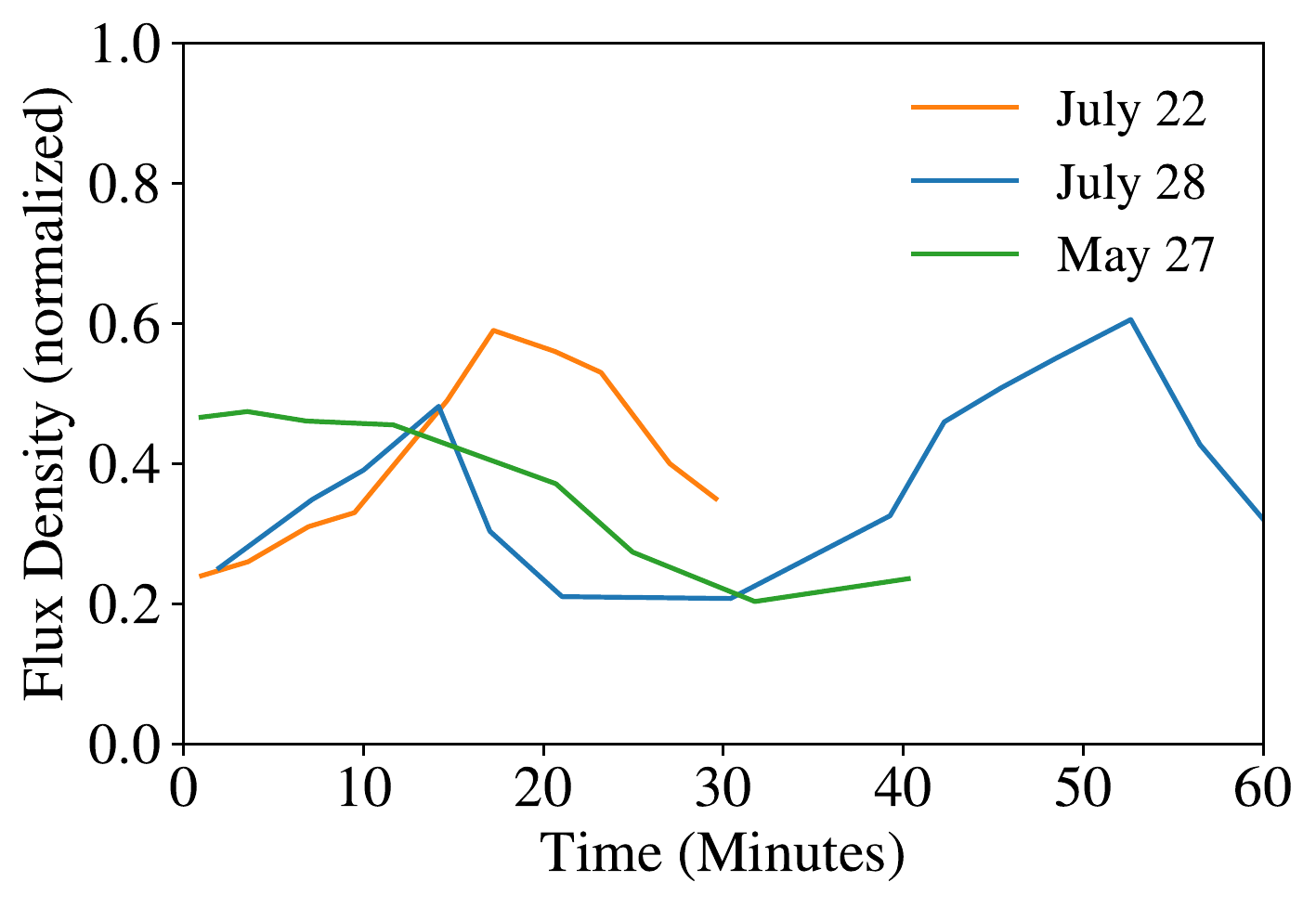}
    \caption{Lightcurves from all three flares observed with GRAVITY.  The May 27th and July 22nd flares show a single peak, while the July 28th flare has a secondary peak separated from the first by $\approx 40$min.}
    \label{fig:lightcurve_data}
\end{figure}

\section{Characteristics of the Flares Observed with Gravity}
In 2018, the GRAVITY collaboration presented high spatial-resolution observations during three flares from \sgra\ and reported that the centroids of the bright spots followed ellipse-like trajectories, with excursions of the order of $\sim 200 \; \mu a s$ (\citealt{gravity2018}). Of the three flares, 
two occurring on July 22nd and July 28th were significantly brighter than the third occurring on May 27th. 

We show in Figure~\ref{fig:centroid_data} the centroid positions from the GRAVITY data for all three observed flares, but with adjusted coordinates compared to those used in the figures in the GRAVITY paper.  In the latter, the coordinates are centered on the median centroid position of a given flare, whereas here we centered the coordinate system on the position of Sgr A* (shown as a black dot at $x=0, \; y=0$) in order to be able to compare their relative locations. It is evident from this figure that none of the orbits are centered on the position of the black hole. Another interesting feature of these centroid tracks is that the trajectory followed by the flare on July 28th (blue dots) appears practically in the polar opposite direction with respect to the black hole (i.e., rotated $180^\circ$) compared to the other two flares, which share a common directionality.

In their original analysis, the GRAVITY paper invoked a hot spot model orbiting in the equatorial plane around the black hole to explain these observations. It attributed this offset between the center of the centroid motion and position of the black hole to the fact that the observations may span only $50-70$ percent of the orbit. In the follow-up analysis, \citet{baubock2020} also considered non-equatorial orbits. Although finding models that better fit the trajectories, the solutions for the different flare orbits presented in the follow-up paper still did not show a common center. 

In addition to the trajectories of the bright spots, GRAVITY also reported on the evolution of the total observed flux over the course of the flares.  We show the combined lightcurves in Figure~\ref{fig:lightcurve_data}. For the May 27th and July 22nd flares, the lightcurves show relatively little structure, falling off monotonically after the first peak. The July 28th flare, on the other hand, shows two distinct peaks in flux that are separated by about $\sim 40$ minutes, which is comparable to the doubly-peaked lightcurve reported by {\em Chandra} during an X-ray flare in \citet{haggard2019}.  Overall, the properties of the July 28th flare are significantly different than the other two: it appears to be oriented on the opposite side of the black hole from the others and shows a different evolution in its lightcurve.  We note, however, that the total observing time was longer for the July 28th flare than for the other two, so it is possible that there may have been a second peak associated with one of the other flares.  The differences between these flares motivated us to search for a physical model that will naturally explain the offset of the centroid orbits, the difference in orientation between different flares, and the reason why some flares are single-peaked while others are double-peaked.

\section{The Plasmoid Model}
In this section, we discuss the framework needed to calculate the time-dependent emission from a plasmoid that forms during a reconnection event in the low-$\beta$ funnel region of a black hole and its appearance to a distant observer. The observed properties of the such an event will depend both on small scale processes, such as energy injection and cooling, as well as large-scale processes, such as the motion of the plasmoid and the transport of emitted radiation in the spacetime of the black hole. On microphysical scales, reconnection injects a distribution of energetic electrons, the properties of which depend on the plasma conditions. Here, we will adopt injection parameters that are appropriate for the low density, low$-\beta$, and high magnetization conditions that are appropriate for the funnel region of \sgra\ (\citealt{werner2017,ball2018}), where magnetization is defined via the parameter $\sigma=B^2/4\pi \rho c^2$. The electrons then cool by emitting synchrotron radiation. 

On larger scales, the plasmoid is not expected to be stationary but to move in the ambient gravitational and magnetic field that is present in the funnel region, on dynamical timescales that are relevant for its position. In addition to this motion, radiation emitted from the plasmoid will also be affected by light bending as it travels from the region near the black hole to the observer at infinity. Because each of these physical effects play a role in determining the observables in the plasmoid model, we describe in the following subsections our treatment of the trajectory of the plasmoid, the evolution of the particle energies within the plasmoid, and the general relativistic radiative transfer that allow us to calculate lightcurves and observed centroid motions at infinity.  


\subsection{The Plasmoid Motion}
We treat the plasmoid as a compact magnetized structure that contains energetic particles and moves in the gravitational and magnetic fields near the black hole. To this end, we define a simple orbit that can be representative of such a motion, by restricting the trajectory of the plasmoid to be on a conical helix.  The initial conditions are defined by parameters $r_0, \; \phi_0, \; \theta_0$, $v_{r0}$, and $v_{\phi 0}$, where $r$, $\theta$, and $\phi$ represent the usual spherical polar coordinates, centered on the black hole, and the subscript 0 reflects the initial values for these parameters. For simplicity, we use a constant velocity $v_r=v_{r0}$, such that the radial coordinate simply increases in time as $r(t)=r_0+v_r t$.  The plasmoid moves on a surface of constant $\theta$, such that $\theta(t)=\theta_0$.  We then solve for an orbit that conserves the Newtonian angular momentum, i.e.,

\begin{equation}
    \dot{\phi}(t)=\dot{\phi}_0 r_0^2 / r(t)^2.
\end{equation}

Naturally, a plasmoid may move on a more complicated trajectory under gravitational, hydrodynamic, and magnetic forces, which would introduce a larger number of parameters to the model. However, our goal here is to identify the simplest physical model that can approximate a set of likely trajectories with the potential to explain the GRAVITY observations. Because of that, we choose to limit the current scope to the conical helix trajectories defined above. We note that both $v_r>0$ and $v_r<0$ cases are allowed in this setup. The case of $v_r>0$ is applicable to a plasmoid forming in the vicinity of an outflowing jet, which pushes the structure along with the outflow; whereas the case of $v_{r}<0$ represents a scenario where hydromagnetic forces are negligible and the plasmoid falls in towards the black hole. The $v_r=0$ case restricted to the equatorial plane reduces to the traditional hot spot model.  We summarize in Table~\ref{param_table} the orbital parameters we use for the two models we explore in this paper, which we will elaborate on further in sections \ref{posterior_plasmoid} and \ref{anterior_plasmoid}.

\capstartfalse 
\begin{deluxetable}{cccccc}
\centering
\tablewidth{\columnwidth}
\tablecaption{Summary of orbital parameters.}
\tablehead{Model& $r_0$ & $\phi_0$ & $\theta_0$ &$v_{r}$ & $v_{\phi}$}
\startdata
Posterior Plasmoid & 36 & $200^\circ$&$15^\circ$&0.01c & 0.41c \\
Anterior Plasmoid & 50& $0^\circ$ & $165^\circ$ & -0.5c & 0.5c 
\enddata
\label{param_table}
\end{deluxetable}
\capstarttrue

\begin{figure*}
    \centering
    \includegraphics[width=\linewidth]{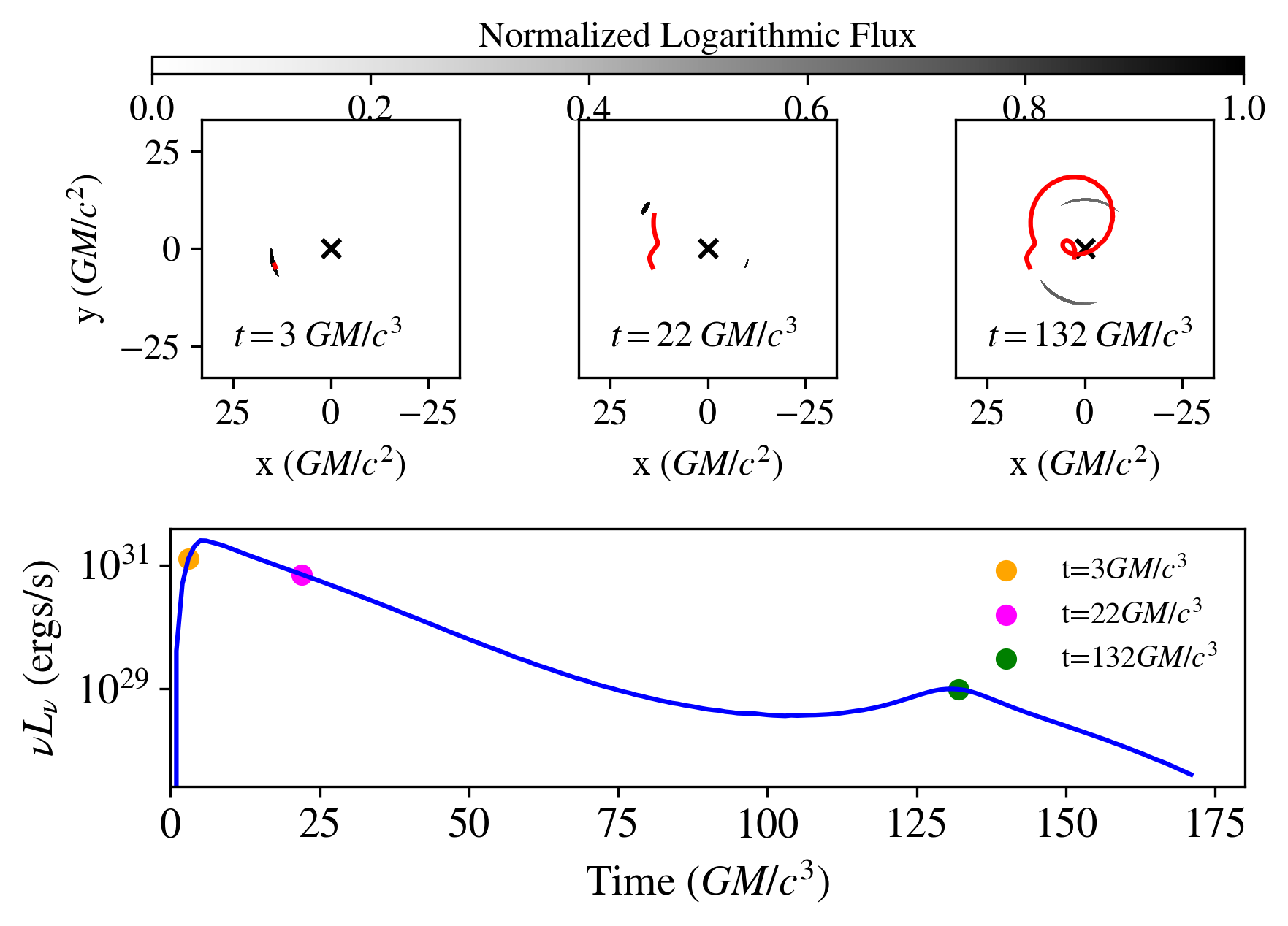}
    \caption{Top: snapshots from three distinct times in a simulation where the plasmoid is in the posterior region of the black hole, ordered chronologically from left to right.  The background color scale from white to black shows the intensity (logarithmically scaled and normalized) at each pixel in the image.  The red line shows the motion of the centroid, up to the time of a given snapshot.  The black cross in the middle shows the location of the black hole.
    Bottom: Lightcurve from the same simulation; the three colored dots indicate the time along the lightcurve that the snapshots in the top panels correspond to. The first peak in the lightcurve appears at $t \approx 4 \; GM/c^3$, at the end of the injection phase. The second peak at $t \approx 132 \; GM/c^3$ is the result of intense gravitational lensing.}
    \label{fig:counter_highres}
\end{figure*}

\subsection{Evolution of the Electron Energy Distribution}
Having specified the orbital motion of the plasmoid, we turn to developing a physically motivated model for the evolution of the electron energy distribution in time throughout the flare. The event begins with an injection phase, where reconnection heats the plasma and loads it into the plasmoid. The injection phase lasts for a time given by $t_{\rm{inj}}$ that is set by the reconnection timescale. At this point, the electrons in the plasmoid begin to cool via synchrotron radiation, which we refer to as the onset of the cooling phase. Even though these phases can in principle overlap, i.e., the cooling can begin before the acceleration phase is over, we choose to treat the two processes as distinct and sequential phases because of the fact that the acceleration timescale is much shorter than the cooling timescale. 


To describe the physics of the injection phase, we define an ``injection energy'', $\gamma_{\rm{inj}}$, which we will take as the typical Lorentz factor the electrons are heated to by reconnection. In general, the heating and acceleration processes generate a distribution of electron energies that depends on the conditions of the plasma (e.g., \citealt{sironi2014,li_guo_2017,werner2018,ball2018}). GRMHD simulations indicate that the magnetization in the jet or corona region of \sgra\ is of order $\sigma=1$ (e.g., \citealt{ball2017}) and a correspondingly low plasma-$\beta$ ($\beta \approx 0.1$) due to the low density of particles in this region. \citet{ball2018} used particle-in-cell simulations to calculate the energization and acceleration of particles in a low-$\beta$ plasma at $\sigma=1$ and found that reconnection heats the peak of the electron distribution to approximately $\gamma=500$. We set this to be equal to $\gamma_{\rm{inj}}$ in this study\footnote{The energy distributions resulting from reconnection also typically contain a power-law tail of higher energy electrons, especially at the lowest values of the plasma $\beta$. We ignore this component here.}.  

In addition to the typical Lorentz factor of the electron at the end of the reconnection event, we also estimate the typical rate at which electrons gain energy during the receonnection event, which we denote as the injection rate $\dot{n}$. The canonical reconnection rate is given by $v_{\rm{rec}}=0.1 c (\sigma/\sigma + 1)^{1/2}$, and the relevant length scale of our problem is the plasmoid size, which we take here to be $L=1 GM/c^2$, consistent with recent observations that infer the size of the emission region (e.g., \citealt{ponti2017}) of flaring electrons.  Using these relations, we can express our injection rate as $\dot{n}=n_0 v_{\rm{rec}}/L$.  For $\sigma=1$, this yields
\begin{equation}
    \dot{n}= \frac{0.07 n_0}{GM/c^3},
\end{equation}
where $n_0$ is the background electron density that flows into the reconnection region and gets energized up to $\gamma_{\rm{inj}}$. For the purposes of this calculation, we take a thermal distribution with a temperature equal to $\theta \approx \bar{\gamma}/3$, i.e., the average Lorentz factor of the electron energy distribution, which is valid in the relativistic limit.  

During the injection phase, the temperature is fixed to $\theta=\gamma_{\rm{inj}}/3$ and the the number density of electrons at a given time, $t$, is simply given by
\begin{equation}
    n=\dot{n}t=\frac{0.07 n_0}{GM/c^3}t.
\label{equation:inj_number}    
\end{equation}
After reconnection and the corresponding particle heating ceases at $t=t_{\rm{inj}}$, the number density of electrons is fixed, and they cool via synchrotron radiation.  The synchrotron power emitted by an electron is (e.g., \citealt{rybicki1979})
\begin{equation}
    P=\frac{4}{3}\sigma_T c \beta^2 \gamma^2 B^{2}/8\pi.
\end{equation}
In the limit of very relativistic electrons (electrons near $\gamma_{\rm{inj}}$ are indeed highly relativistic), this gives
\begin{equation}
    m_e c^2 \dot{\gamma}=\frac{4}{3}\sigma_T c \gamma^2 B^{2}/8\pi.
\end{equation}
We can then write an expression for the cooling rate, $\dot{\gamma}$, as
\begin{equation}
    \dot{\gamma} = 3.2 \times 10^{-6} \times \left(\frac{B}{50}\right)^2 \gamma^2\; s^{-1}.
\label{equation:synch_cooling}
\end{equation}
Solving this equation for the Lorentz factor as a function of time yields
\begin{equation}
    \gamma(t)=\frac{1}{1/\gamma_0 + Ct},
\label{equation:synch_cool}
\end{equation}
where $C=3.2 \times 10^{-6}$ is the coefficient in equation (\ref{equation:synch_cooling}).

\subsection{Computing Observables at Infinity}
We now calculate the lightcurves and trajectories resulting from the emission from the plasmoid as viewed by a distant observer. We implement the motion of the plasmoid as well as the expressions for the number density and the Lorentz factor during injection and cooling phases given in Eq.~(\ref{equation:inj_number}) and Eq.~(\ref{equation:synch_cool}) into a general relativistic radiative transfer simulation. To this end, we use GRay (\citealt{chan2013}), which we have modified to account for the finite speed of light. GRay integrates the radiative transfer equation along geodesics in a black-hole spacetime and includes all of the general-relativistic effects. 

In GRay, we initialize a square grid of $1024 \times 1024$ "rays" over a field of view that we vary depending on the particular problem, set at a distance $1000 GM/c^2$ away from the black hole.  More specifically, we choose a field of view, centered on the black hole, that is just wider than the plasmoid orbit.  In this way, we ensure that the entire motion of the plasmoid falls within the field of view while maximizing the resolution given our number of rays. We integrate each ray backwards along a null geodesic towards the black hole and numerically integrate the radiative transfer equation along these paths. For the results presented in this paper, we perform this calculation for the case of a high-spin Kerr black hole with $a=0.9$, where $a$ is the dimensionless black hole spin parameter $J/M^2$.  We note that the choice of black hole spin here is somewhat arbitrary due to the large uncertainty on measurements of the spin of Sgr~A*.  We did, however, test various values of the black hole spin and found that the salient qualitative features in both the centroid orbits and lightcurves persisted across a wide range of spins.  The precise quantitative details (e.g., time delay, magnitude of secondary peaks, and centroid motion), however, can vary somewhat depending on the spin and geometry of the setup.

\section{Plasmoids in the Posterior Region}
\label{posterior_plasmoid}
Using the setup described in the previous section, we explore the parameter space of plasmoid orbits to identify models that can adequately fit the centroid motion and the lightcurve observed during the July 22nd flare.  In principle, a number of combinations of parameters within our model can reproduce the general circular shape that is apparent in the July 22nd flare orbit. Because we opted to keep the level of complexity of the model and the number of model parameters fairly small, our goal is to identify a class of models that are able to describe the data reasonably well rather than to perform a formal multi-parameter search to find the best-fit orbit. We focus, in particular, on reproducing the following set of salient features: the general characteristics of the centroid motion, the number of re-brightening events in the lightcurve, the time between re-brightening events, and the luminosity ratio between peaks, when more than one peak exists.

As a representative example of a plasmoid orbit that describes the centroid motion of the July 22nd flare reasonably well, we show a model where the plasmoid is in the funnel region on the opposite side of the black hole from the observer, which we refer to as the "posterior plasmoid" model. We use the following parameters for the plasmoid motion: the polar angles are $\phi_0=200 ^\circ$ and $\theta=15^\circ$; the initial distance from the black hole is $r_0=36 \; GMc^{-2}$, and the initial radial and azimuthal components of the velocity are $v_{\phi}=0.41c$, and $v_r=0.01c$. The centroid motion from the July 22nd flare exhibits an orbit with a fairly constant radius, which suggests $v_r  \ll v_{\phi}$. For this reason, we use a value of $v_r \sim 0$ for this particular model.  We set the inclination of the observer to $\theta_{\rm{obs}}=168^\circ$, which is close to the fiducial inclination of $160^{\circ}$ that was used by the GRAVITY collaboration. (Note that this places the observer on the opposite side of the black hole from the plasma, which is oriented at $\theta=15 ^\circ$; hence the designation of this set up as a "posterior plasmoid'' model.) For the local properties of the plasmoid, we set its size equal to $1 \; \rm{GM}/\rm{c}^2$, its magnetic field strength to $35$ G, and the background density, $n_0$, to $10^6 \; cm^{-3}$.  

\begin{figure}[!h]
    \centering
    \includegraphics[width=\linewidth]{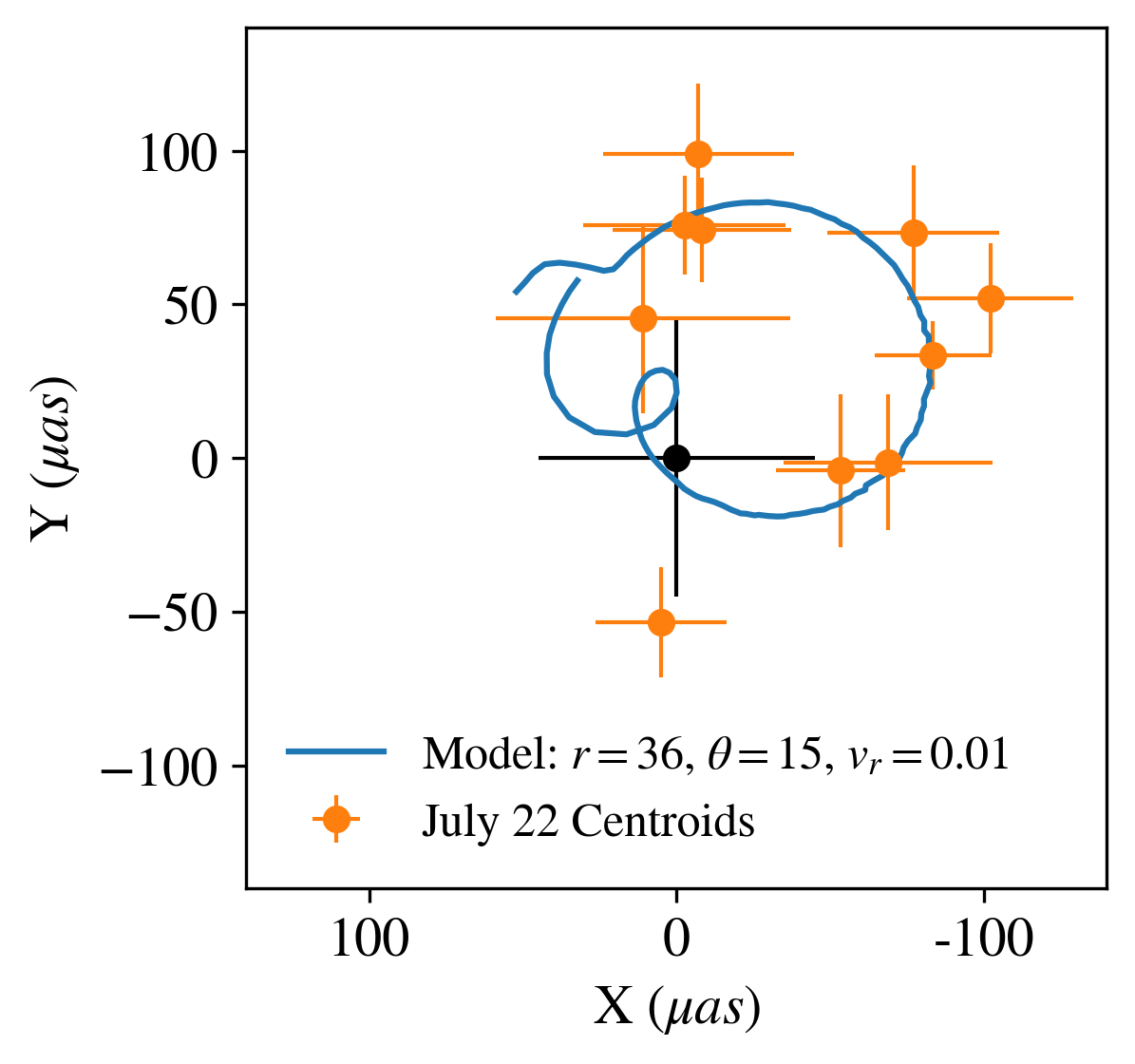}
    \caption{Observed centroid positions from July 22nd flare (orange dots) and centroid track from posterior plasmoid model (blue line).  We see that this model captures the general features from the data, including a mostly circular orbit, offset from the center of the black hole.}
    \label{fig:counter_centroid}
\end{figure}

We show in the top panels of Figure~\ref{fig:counter_highres} the snapshots of the image at three notable times in the simulation.  In these plots, the color scale corresponds to the intensity of light at a given position in the image plane (where each pixel in the image corresponds to a null geodesic optical path, or, "ray"). The red line depicts the centroid motion up to the time in that snapshot, while the black cross shows the location of the black hole.  In the bottom panel, we show the lightcurve with a blue line, and mark the times we show with colored dots along the lightcurve.  

In the first of the snapshots (top left panel), the injection phase has just ended and the light from the secondary image has not yet reached the observing plane.  This is because the path length associated with the secondary image is slightly longer than the path length associated with the primary image, causing a delay in the appearance of the secondary image.  In the second snapshot (top middle panel), a secondary image forms in the bottom right quadrant, roughly mirrored across the black hole from the direct image of the plasmoid in the upper left quadrant.  This secondary image is the result of null geodesics that are strongly lensed by the black hole.  Although the secondary image appears in the second panel, it does not result in a significant flux increase that is discernible in the lightcurve: the flux from the secondary image here is subdominant to the flux from the primary image, but it does result in a slight shift of the centroid inwards.  In the final snapshot (top right), we see a strong lensing event, where both the primary and secondary images are highly elongated.  The lensing causes a significant increase in flux, by a factor of approximately 5.  However, at this point in the simulation, the plasmoid has cooled sufficiently such that, despite the factor of $\sim 5$ increase from the strong lensing, the luminosity of the second peak is $\sim 2$ orders of magnitude smaller than the peak associated with the initial rise of the flare.

We show in Figure~\ref{fig:counter_centroid} the full centroid motion from the posterior plasmoid model with a blue line and the centroid data from the July 22nd flare with orange dots.  We see that the data are reasonably well described by $\sim 75\%$ of the full orbit from our model.  This is similar to the GRAVITY interpretation that the centroid motion associated with the July 22nd flare can be explained by a partial orbit.  Our model differs, however, because it naturally accounts for the offset between the black hole and the center of the centroid motion.

\begin{figure}[!h]
    \centering
    \includegraphics[width=\linewidth]{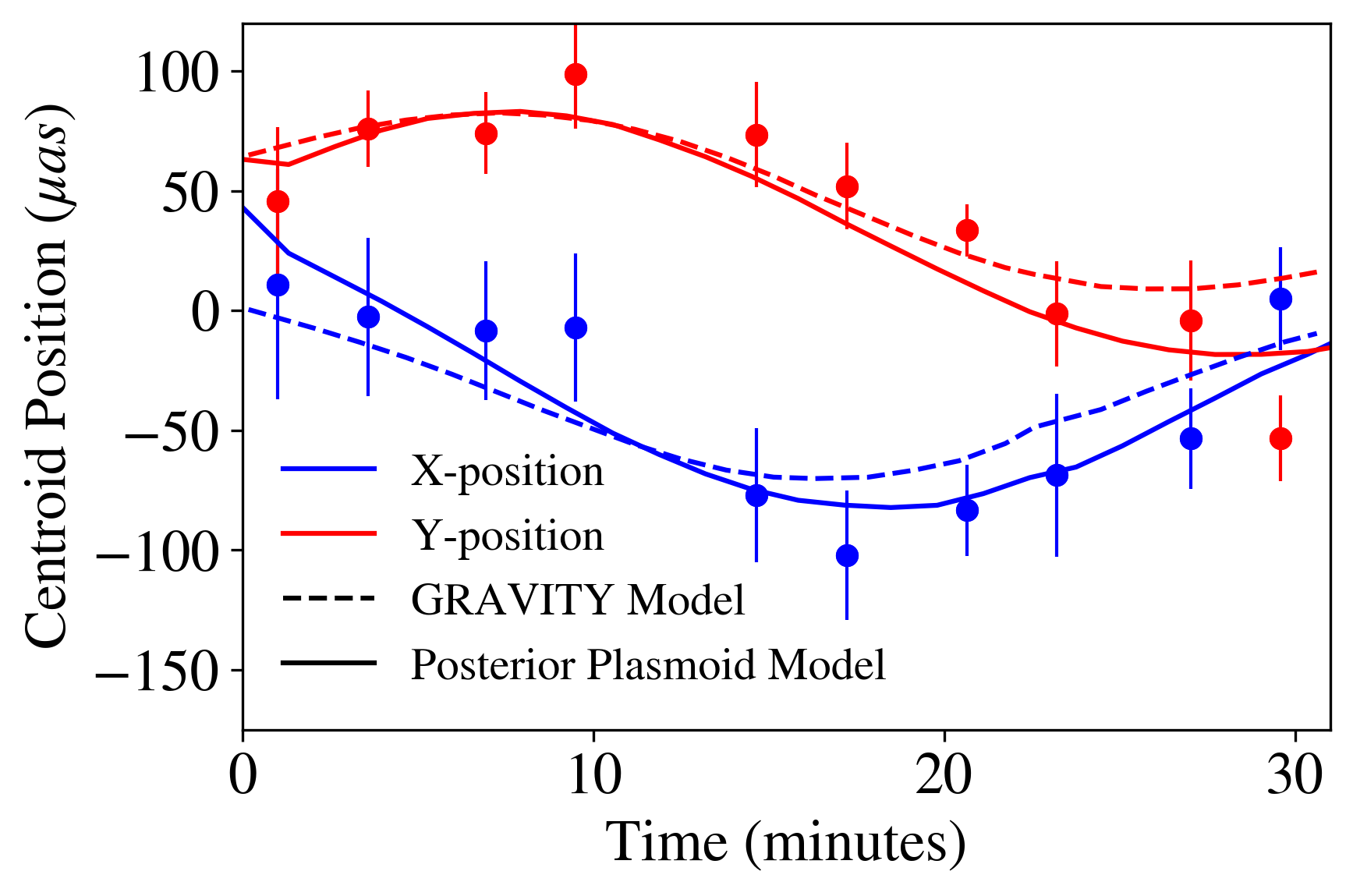} 
    \caption{X (blue) and Y (red) centroid positions from our posterior plasmoid model (solid lines), the model from the GRAVITY paper (dashed lines), and the observed centroid positions (points). Both models fit the data equally well. }
    \label{fig:model_fit}
\end{figure}

In order to see how our model compares to the orbital model that was proposed by the GRAVITY collaboration to interpret the observations (i.e., an orbit in the equatorial plane with a radius of $7 GM/c^2$, viewed at an inclination of $160^\circ$), we  show in Figure~\ref{fig:model_fit} the X (blue) and Y (red) centroid positions from the posterior plasmoid model described above (solid lines), the model shown in the GRAVITY paper (dashed lines), as well as the observed centroid positions (points). With the caveat that neither study employed formal model fitting to estimate the best parameters, we find that the plasmoid model provides a similar description of the data as the orbital model. 

\subsection{Phenomenology of other Plasmoid Orbits}
\begin{figure}[!h]
    \centering
    \includegraphics[width=\linewidth]{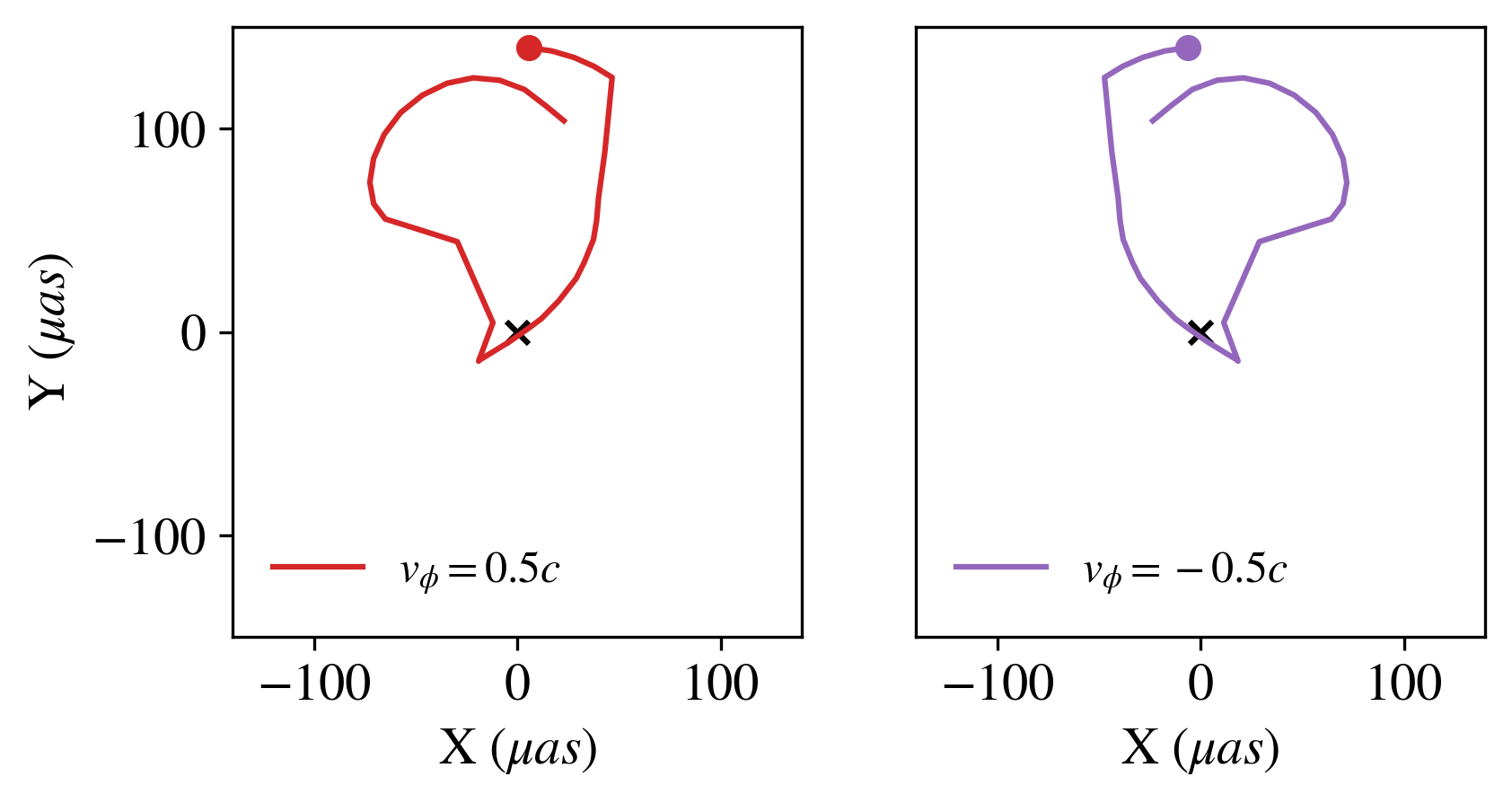} 
    \caption{Two identical posterior plasmiod models, but with the direction of plasmoid motion reversed (left: $v_{\phi}=0.5c$, right: $v_{\phi}=-0.5c$).  We denote the starting position of the plasmoid with a filled-in dot.  We see that by reversing the direction of motion of the plasmoid, that the shape of the centroid orbit fundamentally changes.}
    \label{fig:many_track}
\end{figure}

When the plasmoid orbit is in the posterior region, the interplay of cooling, finite light travel time, and lensing results in a rich variety of possible perceived centroid motions.  For instance, in Figure \ref{fig:many_track}, we show two models, both with an initial radial location $r_0=40 GM/c^2$, radial velocity $v_{\rm r}=0.01 c$, and observed at an inclination of $i=160^{\circ}$. We denote the starting positions with filled-in dots.  The difference between the two models is the direction of the plasmoid orbit: the left panel shows the centroid motion of a plasmoid moving at $v_{\phi}=0.5c$ and the right panel shows a nearly identical model but with $v_{\phi}=-0.5c$ and an associated black hole spin flipped (a=-0.9), such that the plasmoid is always corotating with the spacetime.  Both of these models exhibit a distinct warp in the observed trajectory, displaying a teardrop-like shape.  The latter is caused when the plasmoid passes closely behind the black hole with respect to the observer's line of sight and is strongly lensed.  The light that forms the lensed image originates from an earlier time, when the plasmoid was hotter, and hence can pull the position of the centroid substantially away from the primary image, resulting in a teardrop-like shape.

Interestingly, by changing the direction of travel of the plasmoid, the perceived centroid motion is mirrored.  Initially, one may expect that reversing the direction of the plasmoid orbit to result in the same shape of the centroid motion, but traveling in the opposite direction.  Instead, we demonstrate here that the finite light travel time coupled to the cooling of the plasmoid breaks the time symmetry such that reversing the direction of the plasmoid motion results in a differently shaped centroid trajectory (or, in this case, a mirrored trajectory).

\section{Plasmoids in the Anterior Region}
\label{anterior_plasmoid}
We now explore the features of the model when the plasmoid is on the same side of the black hole as the observer, which we refer to as the "anterior plasmoid'' model.  This is motivated not only by a desire to explore the full range of phenomena that can occur in the simulations but also by the fact that the trajectory of the July 28 flare occurs in the opposite quadrant with respect to the black hole than do the July 28 and May 27 flares. 

We use the same plasmoid properties (i.e., magnetic field strength, size, and density) for this model, but employ slightly different orbital parameters to highlight some of the most interesting phenomena that can occur in the anterior setup.  For the orbital parameters, we use a radial distance $r_0=50$, polar angles $\phi_0=0^\circ$ and  $\theta=165^\circ$, and azimuthal and radial velocities $v_\phi=0.5c$ and $v_r=-0.5c$.  There are two main differences between this model and the posterior model. First, as previously mentioned, is that the plasmoid is on the same side of the black hole as the observer (the observer's inclination is set to $\theta_{\rm{obs}}=165^\circ$ for all models). Second, the plasmoid has the opposite sign of radial velocity and is falling in towards the black hole, which as we will see below, can result in strong double peaks in the lightcurve.

\begin{figure*}
    \centering
    \includegraphics[width=\linewidth]{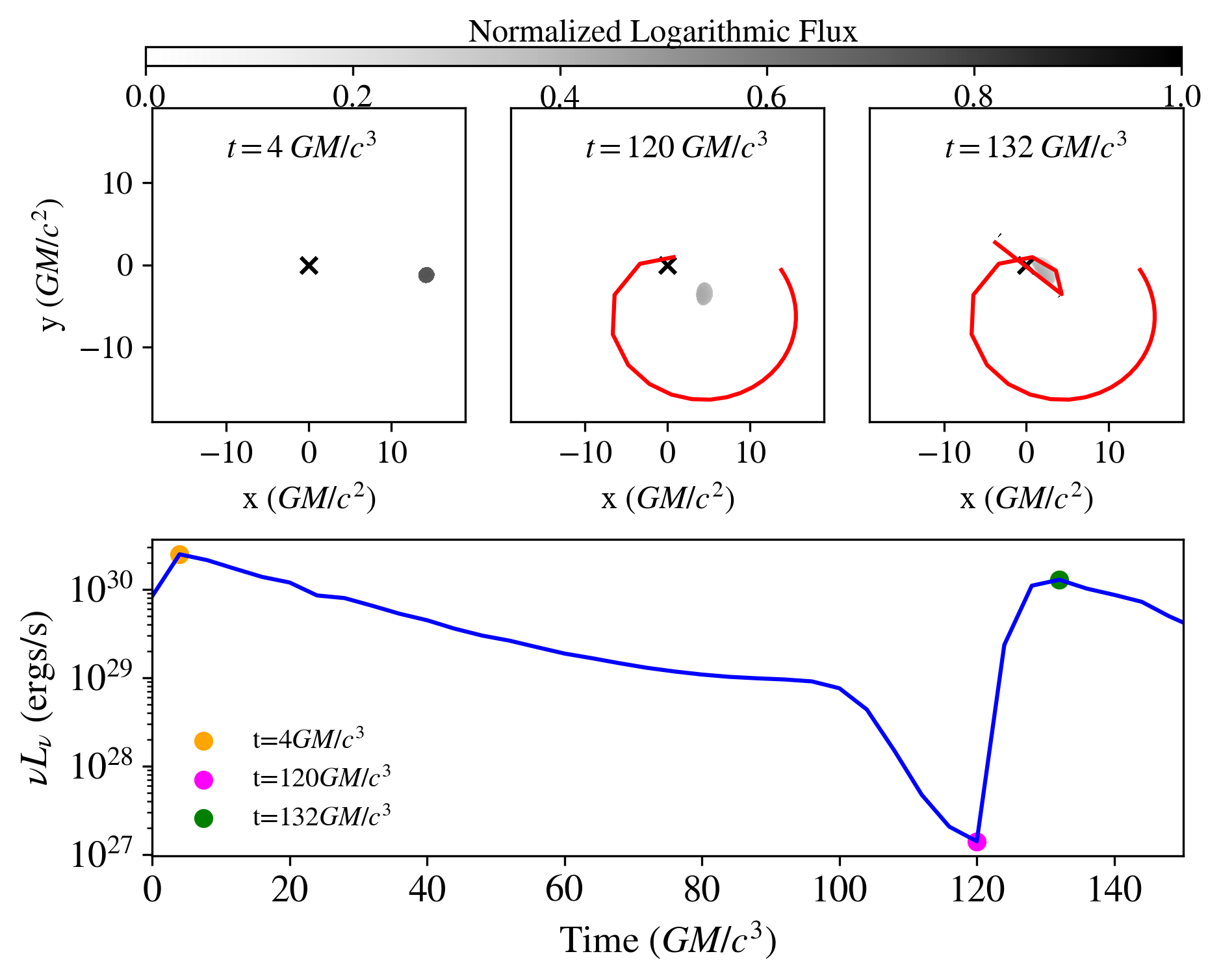}
    \caption{Top: snapshots from three distinct times in the anterior plasmoid simulation, increasing chronologically from left to right.  The background color scale from white to black shows the logarithmically scaled and normalized intensity of the image.  The red line shows the motion of the centroid, up to the time of a given snapshot.  The black cross in the middle shows the location of the black hole.
    Bottom: Lightcurve from the simulation, with the three colored dots showing the time along the lightcurve that the snapshots in the top panels correspond to. There is a second peak at $t=132 GM/c^3$, accompanied by a sudden change in centroid position, similar to that seen in the July 28 flare.  This is caused by light emitted at the  beginning of the simulation, but has traveled around the back of the black hole to reach the observer.  Because this light is blue shifted due to the motion of the centroid and is emitted at a time before the plasmoid had significantly cooled, it results in a sudden increase in flux.  The solid angle associated with this secondary image, however, is small, resulting in only a few pixels capturing the extent of the secondary image.}
    \label{fig:forward_highres}
\end{figure*}

We show in Figure~\ref{fig:forward_highres} an analogous plot to Figure~\ref{fig:counter_highres}, but for the anterior setup.  We see that the secondary image does not occur until very late times (middle panel, corresponding to $t\approx 132 GM/c^3$ in this model), when the centroid shifts rapidly and dramatically, as the delayed light from the secondary image hits the observer's plane. In order to form the secondary image in the anterior plasmoid model, the light must travel from the plasmoid towards the black hole (away from the observer) a distance $r_0$, then an additional distance $\sim 15 GM/c^2$ around the black hole in the vicinity of the photon ring such that the photon direction changes substantially ($\sim 180^\circ$), and then back again to the observer, another $r_0$ in distance.  We note here that we are using the term photon ring loosely to describe spherical photon trajectories that exist in the vicinity of Kerr black holes; these trajectories are rings only for the case of non-spinning black holes. In either case, the locations of the spherical photon orbits provide a useful mark for the region in the spacetime at which the photon trajectory can change direction by $\sim 180^\circ$.   Because of this extra distance traveled, the light that the observer sees from the secondary image is delayed with respect to light emitted at the same time that travels directly to the observer by $\approx 2r_0/c+15 GM/c^3$. We see a strong second peak occur at this time (also see last panel at the top): when light from the delayed lensed image reaches the observer's plane, it shifts the centroid position so rapidly that it appears as a superluminal motion and also results in a strong second peak in the lightcurve.  The time difference between peaks is comparable to the time between the two peaks in flux observed by the GRAVITY collaboration during the July 28th flare, which is $\approx 40$ minutes, or 120 $GM/c^3$). 

It is noteworthy that the simple set up that explains the particular observed trajectories of the flares with respect to the position of the black hole also naturally produces the differences observed in the lightcurves between the flares. Because the double-peaked lightcurves is an unusual characteristic of \sgra\ flares that are observed both with GRAVITY and {\it Chandra}, we turn to a more thorough exploration of this phenomenon in the next section. 

\section{The Appearance of Double Peaked Lightcurves}
We showed that an anterior plasmoid model with an initial plasmoid distance from the black hole of $r=50\; GM/c^2$ naturally results in a distinct double-peaked lightcurve, with a time between the two peaks that matches the time between observed double peaked lightcurves in GRAVITY and {\it Chandra} observations.  In this section, we further explore how the properties of these two peaks depend on the orbital parameters of the plasmoid.  First, we focus on how the infall speed, $v_r$, influences the relative amplitude of the two peaks.  In the context of the anterior plasmoid model, we expect a larger infall speed to cause the direct image of the plasmoid to be redshifted and dimmed, while the secondary image is boosted.  

\begin{figure}[!h]
    \centering
    \includegraphics[width=\linewidth]{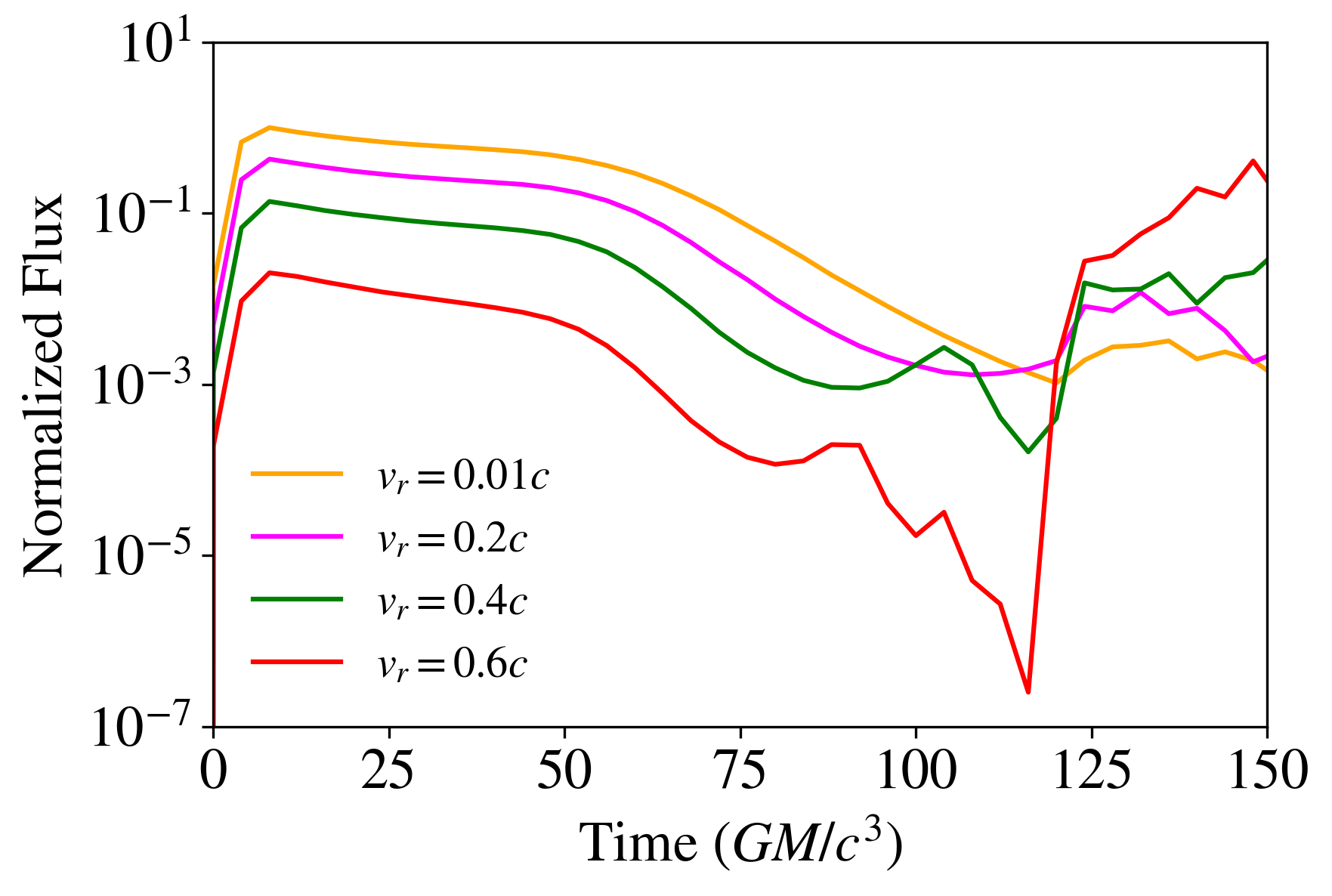}
    \caption{Lightcurves from a number of anterior plasmoid models holding all parameters fixed while varying the radial velocity $v_r$ (defined as pointing inward, towards the black hole). Increasing the infall velocity has the effect of redshifting (and hence dimming) the direct image (first peak), while blueshifting (boosting) the light from the secondary image (second peak).  The secondary image appears at a significantly delayed time at $\sim 2 r_0$, the light travel time corresponding to the initial distance of the plasmoid. The relative height of the two main peaks is set by the infall speed, with the expected general behavior.}
    \label{fig:vr_lightcurves}
\end{figure}

We show in Figure~\ref{fig:vr_lightcurves} the lightcurves from four anterior plasmoid simulations, with varying values of the radial velocity $v_r$, while all other model parameters are held constant.  Each lightcurve is normalized to the maximum flux from the $v_{r}=0$ model. We see that all of the models show a distinct second bump that starts at roughly $t=120 \; GM/c^3$.  The amplitude of this second peak, however, depends strongly on $v_r$, with larger infall speeds causing a brighter second peak.  Additionally, we see that the amplitude of the first peak also depends on $v_r$, but with the opposite trend of the second bump. This is easy to understand in the context of redshifts and blueshifts of the direct and secondary images.  The first peak originates from light emitted towards the observer, forming the "direct'' image.  As the infall speed $v_r$ increases, the plasmoid is moving away from the observer faster, resulting in a redshift and dimming of the light that the observer receives directly from the plasmoid.  The second peak, however, is dominated by light that is emitted in the direction of the plasmoid motion, orbits around the black hole, and then reaches the observer's plane.  The solid angle associated with this secondary image will be very small relative to the direct image, which is why the amplitude of the second peak is much smaller when the plasmoid is not moving ($v_r=0$).  However, as the infall speed increases, we see that the amplitude of the second peak rises, and eventually becomes greater than the amplitude of the first peak, as in the $v_r=0.6 \; c$ case.  This occurs because the plasmoid is blueshifted along the line of sight that forms the secondary lensed image.

\subsection{Decomposing the Effects of Boosting, Lensing, and Gravitational Redshift}
The lightcurves shown in Figure~\ref{fig:vr_lightcurves} reveal a rich structure, beginning with an initial peak and relatively slow cooling, followed by a faster decline with numerous smaller rises and falls in the lightcurve, and ending with a strong second peak. We now aim to delineate the physical mechanisms responsible for the dominant features.  To this end, we run a simplified set of models, where we systematically and artificially exclude various pieces of physics and explore how these choices affect the resulting structures in the lightcurve.

\begin{figure}[!h]
    \centering
    \includegraphics[width=\linewidth]{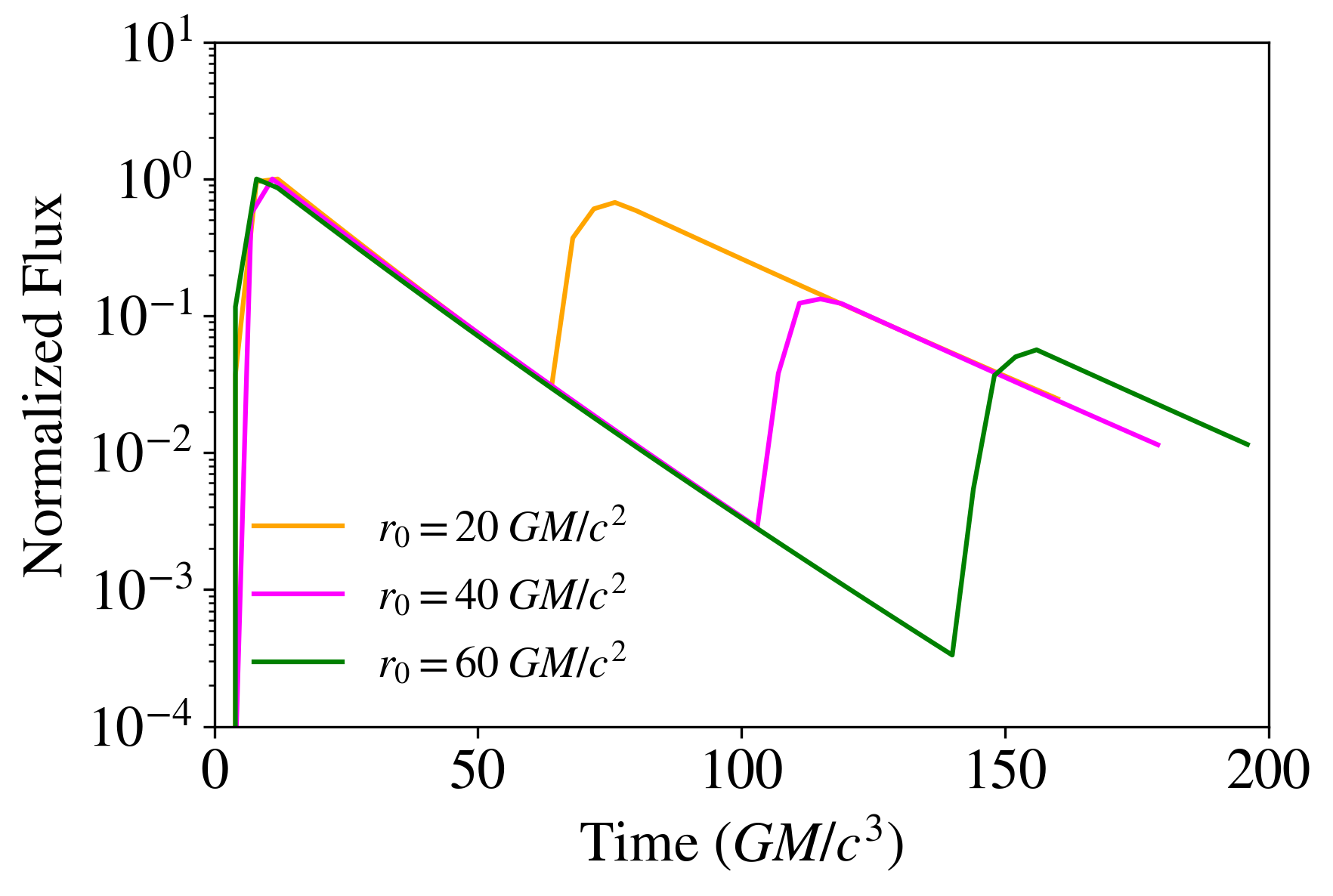}
    \caption{Lightcurves from a number of anterior plasmoid models with no plasmoid motion.  By removing additional effects from plasmoid motion, we cleanly isolate the effect of finite light travel time.  We clearly see the expected behavior of both the timing of the second peak as well as its relative brightness to the primary peak.  As the plasmoid gets further from the black hole, the time to the secondary peak is delayed, and the secondary peak also becomes dimmer because the solid angle subtended by the secondary image becomes smaller, as fewer optical paths intercept the plasmoid.}
    \label{fig:r0_lightcurves}
\end{figure}

First, we aim to cleanly isolate the cause of the strong double-peak, without the confounding effects of any additional features in the lightcurve. We have compelling reasons to believe that this delayed peak is caused by the delayed lensed image of the plasmoid, because {\it (i)} the second peak has the expected $v_r$ dependence, {\it (ii)} the time difference between the peaks matches the difference in the light travel time between the direct and secondary images, and {\it (iii)} the sudden centroid motion is coincident with the appearance of the second peak.  However, we further test this interpretation here by removing the effects of the changing position of the plasmoid.  We show in Figure~\ref{fig:r0_lightcurves} lightcurves from a set of models with a stationary plasmoid, with a varying initial plasmoid distance $r_0$ from the black hole, in order to cause different amounts of time delay.  Due to the lack of motion, changes in the gravitational redshift and lensing as the plasmoid moves will not occur.  Additionally, because the azimuthal velocity is set to 0, there is no Doppler boosting or dimming on orbital timescales. However, in order to achieve a distinct and large second peak, we artificially incorporate a Doppler boost corresponding to $v_r = -0.5 \; c$ in the radiative transfer calculation, even though the plasmoid does not move. We emphasize that this is not meant to be a physical model, but we use it to cleanly show the effects of cooling and finite light travel time without the confounding effects due to the changing position of the plasmoid.  

We see in Figure~\ref{fig:r0_lightcurves} the general features that we would expect from varying the radial distance of the plasmoid from the black hole.  In particular, the time between the first and the second peak, $\Delta t_{}$, increases by $2\Delta r_0$ as $r_0$ increases.  This particular scaling happens simply due to the geometry of the setup: the secondary image is formed by light that travels from the plasmoid towards the black hole, wraps around the black hole, and then comes back towards the plasmoid, ultimately hitting the observer's plane, i.e., it travels the space between the plasmoid and the black hole twice.

\begin{figure}[!h]
    \centering
    \includegraphics[width=\linewidth]{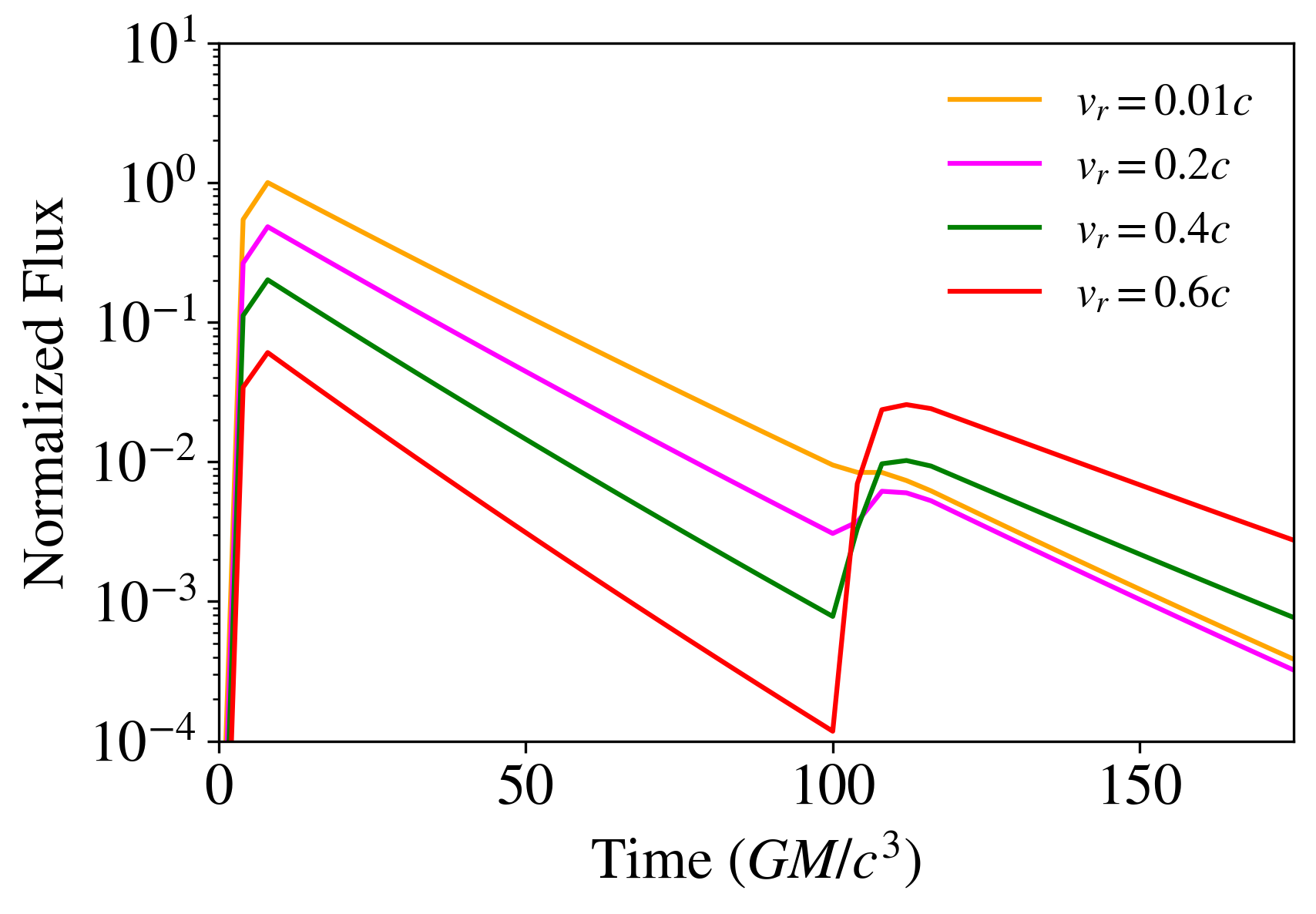}
    \caption{Lightcurves from a set of anterior plasmoid models with no plasmoid motion.  Here we vary the radial velocity used in the radiative-transfer portion of the calculation, despite the plasmoid not physically moving throughout the simulation, to isolate the effect of Doppler boosting on the properties of the double peaked behavior in the lightcurve.}
    \label{fig:vr_stationary_lightcurves}
\end{figure}

In order to further understand the simple properties of the double peak without other physical effects, we show in Figure~\ref{fig:vr_stationary_lightcurves} lightcurves from a set of simulations with a stationary plasmoid, but varying the radial velocity used in the radiative transfer equation.  We see that we cleanly recover the distinct double-peaked behavior and underlying trends we found in Figure~\ref{fig:vr_lightcurves} that included all of the physics resulting from plasmoid motion.  Specifically, we see that when the plasmoid's inward velocity is higher, the first peak corresponding to the direct image is dimmed while the secondary lensed image is boosted.  By comparing the lightcurves in Figure~\ref{fig:vr_lightcurves} and \ref{fig:vr_stationary_lightcurves}, we can identify the additional effects of plasmoid motion.  As the plasmoid falls into the gravitational well of the black hole, the gravitational redshift dims the light the observer receives from the plasmoid at a rate that becomes faster than the cooling, which explains the sudden dip in the lightcurves in Figure~\ref{fig:vr_lightcurves} that sets in faster for models with a higher infall velocity.  Finally, we see a number of smaller peaks and dips in the lightcurve in Figure~\ref{fig:vr_lightcurves} that are not present in Figure~\ref{fig:vr_stationary_lightcurves}.  These occur due to the effects of Doppler boosting/dimming on orbital timescales, and hence are not present in the model shown in Figure~\ref{fig:vr_stationary_lightcurves} where the plasmoid has no azimuthal velocity.

\section{Predictions and Future Outlook}
In this paper, we presented a plasmoid model that explains the offset between the location of Sgr~A* and the centroid orbits during IR flares, their orientation relative to each other, and how one preferred direction (the anterior plasmoid model) naturally produces secondary peaks in the lightcurve if the plasmoid falls toward the black hole.  From this model, we can make a number of predictions for future observations regarding the relationship between the structures in the lightcurve and orientation of the centroid motion during flares.  

First, this model predicts that future observations will continue to reveal centroid orbits with centers that are offset from the position of the black hole.  In this interpretation, the offset is caused by the physical location of the orbit and is not just a consequence of observing only part of the orbit. Second, this model predicts that the centroid positions will orbit along the two sides of a preferred axis, which correspond to the plasmoid being either above or below the black hole and relatively close to the spin axis.  In other words, we expect that future observations will see centroids that either align with the directionality of the May 27th and July 22nd centroid positions (up and to the right in Figure~\ref{fig:centroid_data}), or along the opposite orientation corresponding to the July 28th centroids (down and to the left in Figure~\ref{fig:centroid_data}), but not in between (up and to the left, or down and to the right).

Third, we predict that double-peaked flares with separations between the two peaks on timescales of $\sim 40$ minutes should occur frequently for flares with the same orientation as the July 28th flare, but not for flares with the same orientation as the May 27th and July 22nd flares.  This is not to say that double peaks are impossible in the framework of a posterior plasmoid setup (in the direction of May 27th and July 22nd flares): we already saw that a second peak can form due to the strong gravitational lensing during the plasmoids orbit.  This second peak, however, will be small unless it occurs shortly after the formation of the plasmoid, before it has significantly cooled.  Because of that, any double-peaked feature that may arise in a posterior plasmoid model (and hence, we predict, in the same direction as the May 27th and July 22nd flares) will be on much shorter timescales than the observed $\sim 40$ minute delay between peaks for the July 28th flare.  Furthermore, within this model, the spectrum during the second peak should look like a blue-shifted version of the spectrum during the first peak.
\\
\\
\\
\\
\section{Conclusions}

In the tenuous and low-$\beta$ plasma regions present in the inner accretion flow surrounding Sgr A*, reconnection events leading to short-lived particle acceleration episodes can be commonly expected.  Earlier work had shown that such events are likely to be associated with flares and impart on these flares characteristics that are unique to the motion of plasmoids close to black holes.  The GRAVITY events resolved the positions of bright regions for the first time during flaring activity, providing an opportunity to see if plasmoid motions, coupled with the effects of GR in the vicinity of the black hole, provide a natural explanation for the properties that have been observed.  

In this paper, we showed that a plasmoid model in the funnel region of a black hole can reproduce some of the important features of the GRAVITY observations and explains why there may be a preferred axis along which these flares are oriented and why the center of their motion is offset from the black hole.  Additionally, we show that strong double-peaked flares are a generic consequence of a blob of plasma falling in towards a black hole when it is oriented on the same side of the black hole as the observer.  In this setup, we show that the second peak in the lightcurve comes from the delayed lensed image and occurs at a time roughly $2r_0 + 15 \; GM/c^3$ after the first peak, and that the relative strength of this second peak is amplified when the infall velocity is higher due to the relativistic Doppler boosting.  We also make a number of predictions for future high-resolution observations of IR flares from Sgr A* that will provide a thorough test of this model.



The centroid positions of the emission during the three GRAVITY flares define an axis, which we identify here with the spin axis of the black hole (or more precisely, in case the black hole is not spinning, with the angular
momentum axis of the accretion flow). Within our model, this axis is oriented approximately 135 degrees East of North (or equivalently 135+180=315 degrees East of North). This orientation is broadly consistent with the inferred angular momentum axis of a cold disk around \sgra\ recently detected by ALMA at much larger distances than those probed by GRAVITY ($\simeq 20,000$ Schwarzschild radii; \citealt{Murchikova2019}). 

Observations of Sgr A* with the Event Horizon Telescope offer the possibility of inferring the orientation of the angular momentum axis of the inner accretion flow by measuring the Doppler-induced asymmetry in the brightness of the emission surrounding the black hole, as was done for the case of M87~\cite{PaperI}. Modeling early EHT observations of Sgr A* with semi-analytic~\cite{Broderick2011, Broderick2016} and GRMHD simulations of accretion flows~\cite{Dexter2012} resulted in orientations of 156$^{+10}_{-17}$~degrees and 160$^{+15}_{-86}$~degrees, respectively, which are consistent with the orientation we infer here. 

Besides inferring the orientation of the angular momentum axis in Sgr~A*, EHT observations may also be able to identify whether significant morphological changes in the inner accretion flow are associated with an increase in emission during flares. This will constrain the possible flaring mechanisms, such as gravitational lensing, Doppler boosting of hot spots, and other such processes that will leave a distinct imprint on the image of the inner 
accretion flow.
\\
\\

\acknowledgements
We thank Lorenzo Sironi for useful discussions.
We gratefully acknowledge support for this work from NSF AST-1715061, Chandra Award No. TM6- 17006X, NASA ATP 80NSSC20K0521, and NSF PIRE 1743747.


\bibliographystyle{apj}
\bibliography{david_bib}
\end{document}